\documentclass[aps,floatfix,twocolumn,letterpaper,prl,superscriptaddress,showpacs]{revtex4}

\usepackage{amsmath}
\usepackage{amsfonts}
\usepackage{amssymb}
\usepackage{mathrsfs}
\usepackage{graphicx}
\usepackage{mathtools}

\usepackage{color}

\DeclarePairedDelimiter{\ceil}{\lceil}{\rceil}
\DeclarePairedDelimiter{\floor}{\lfloor}{\rfloor}

\newcommand{\defeq}{\mathrel{\mathop:}=}

\DeclareMathOperator{\fd}{F_D}
\DeclareMathOperator{\herm}{H}

\begin{document}

\title{Quenching to unitarity: Quantum dynamics in a 3D Bose gas}
\author{A. G. Sykes}
\affiliation{JILA, University of Colorado and National Institute of Standards and Technology, Boulder, Colorado 80309-0440, USA}
\author{J. P. Corson}
\affiliation{JILA, University of Colorado and National Institute of Standards and Technology, Boulder, Colorado 80309-0440, USA}
\author{J. P. D'Incao}
\affiliation{JILA, University of Colorado and National Institute of Standards and Technology, Boulder, Colorado 80309-0440, USA}
\author{A. P. Koller}
\affiliation{JILA, University of Colorado and National Institute of Standards and Technology, Boulder, Colorado 80309-0440, USA}
\author{C. H. Greene}
\affiliation{Dept. of Physics, Purdue University, West Lafayette, Indiana 47907-2036, USA}
\author{A. M. Rey}
\affiliation{JILA, University of Colorado and National Institute of Standards and Technology, Boulder, Colorado 80309-0440, USA}
\author{K. R. A. Hazzard}
\affiliation{JILA, University of Colorado and National Institute of Standards and Technology, Boulder, Colorado 80309-0440, USA}
\author{J. L. Bohn}
\affiliation{JILA, University of Colorado and National Institute of Standards and Technology, Boulder, Colorado 80309-0440, USA}

\begin{abstract}
 We study the dynamics of a dilute Bose gas at zero temperature following a sudden quench of the scattering length from a noninteracting Bose condensate to unitarity (infinite scattering length). We apply three complementary approaches to understand the momentum distribution and loss rates.  First, using a time-dependent variational ansatz for the many-body state, we calculate the dynamics of the momentum distribution. Second, we demonstrate that, at short times and large momenta compared to those set by the density, the physics can be well understood within a simple, analytic two-body model. We derive a quantitative prediction for the evolution of Tan's contact, which increases linearly at short times. We also study the three-body losses at finite densities.
 Consistent with experiments, we observe lifetimes which are long compared to the dynamics of large momentum modes. 
\end{abstract}

\maketitle

Ultracold atomic physics offers unique opportunities to study strongly correlated systems due to the tunability of the interatomic interaction, parameterized by the $s$-wave scattering length, $a$, via Fano-Feshbach resonances~\cite{FeshbachResonanceReview}.  Particularly interesting are quantum gases at unitarity, where $a$ is much larger than any other length scale in the system.  Such systems are predicted to exhibit universal behaviour, which depends only on the mean interparticle separation.  Here the physics is highly non-perturbative, with no obvious small parameter.  Investigations to date have predominantly focused on the Fermi gas, where three-body recombination is naturally suppressed by statistical repulsion~\cite{PetrovDimersOfFermionicAtoms}. Over the last decade, a general consensus seems to have emerged on many issues surrounding the unitary Fermi gas~\cite{JohnThomasScience2005,Jin40K-PRL2006,UedaUnitaryFermiGas,SalomonUniversalFermiGas,ZwierleinScience2012}. Theoretical understanding of the unitary Bose gas is far less developed. Although experiments in the quantum degenerate regime have been able to measure beyond mean field effects, such as the famous Lee-Huang-Yang correction~\cite{JinCornellLHYCorrection2008,SalomonLHYMeasurementStronglyInteracting} for values of $na^3\lesssim7\times10^{-3}$ ($n$ being the three dimensional number density), progress towards a unitary Bose gas with $na^3\gg1$ is hampered by the catastrophic scaling of three-body loss in the system. At zero temperature, in the dilute gas, $na^3\ll 1$, with $a\gg r_0$ where $r_0$ is the van der Waals length, the three-body recombination constant scales universally as $L_{3}\propto \hbar a^4/m$~\cite{FedichevShlyapnikovThreeBodyLoss,NielsenMacekThreeBodyRecombination1999,EsryGreeneThreeBodyLoss,BraatenRecombinationLargeScatteringLength,BraatenReview}. This $a^4$ scaling renders any adiabatic transfer from the weakly interacting limit to the unitary limit impossible~\cite{WiemannStronglyInteractingBEC2002,GrimmThreeBodyRecombinationScaling}. As $a\rightarrow\infty$, although remaining dilute compared to the van der Waals length, the long-range aspects of Efimov physics become important~\cite{NielsenMacekThreeBodyRecombination1999,EsryGreeneThreeBodyLoss,BraatenRecombinationLargeScatteringLength,BraatenReview}. One approach to limiting losses has emerged by considering non-degenerate unitary Bose  gases~\cite{SalomonLifetimeWithResonantInteractionsFiniteTemp,HadzibabicStabilityUnitaryBoseGasFiniteTemp2013}, where the thermal de Broglie wavelength, $\lambda_{\rm T}$, can provide a small parameter $n\lambda_{\rm T}^3\ll1$, and {\it low-recombination} regimes exist~\cite{LiHoUnitaryBoseGasFiniteTemp2012}.

A brazen new approach adopted in a recent experiment~\cite{PhilDebbieEricUnitaryBoseGas} utilizes an effectively diabatic quench in the scattering length to unitarity, with the initial gas temperature deeply degenerate. Although dimensional analysis requires both the loss rate and the equilibration rate to scale as $n^{2/3}$,  fortune favoured the brave, and they observed the formation of local (quasi-) equilibrium over faster timescales than the decay.  This exciting result indicates a unique possible route to the experimental realization of a meta-stable unitary Bose gas. However, such a rapid quench toward unitarity places theory in unfamiliar territory and raises several questions. The system is in a dynamic state, neither zero temperature, nor finite temperature. What are the dynamical properties of this state post quench? Can we understand the prevalence of the equilibration rate over the three-body loss rate in this scenario? Over what time scale does an equilibrium state form, and what is the nature of this state?

\begin{figure}[t!]
 \includegraphics[width=8cm]{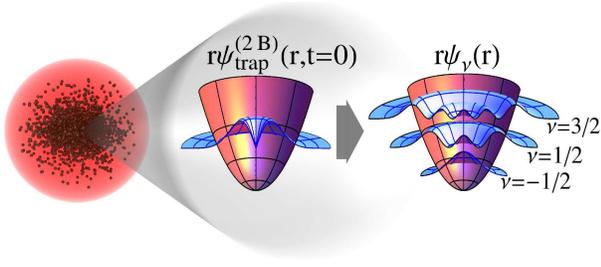}
\caption{(Color online) In the period of time immediately following a quench in the scattering length, the dynamics originates in {\it causally isolated} regions of the cloud. The collective effect of other particles in the system is modelled by an artificial trap~\cite{GreeneArtificialTrapsForTwoBodySystemsNJP2003, PaulJulienneTwoBodyArtificialTrapsJPhysB2004}, tailored to give the particles a mean-interparticle-separation which is consistent with that of the many-body system.}
\label{fig-cartoon}
\end{figure}

In this letter we study and compare the results from several different models to understand the dynamics immediately following a quench from noninteracting ($a\lesssim r_0$) to unitarity ($a=\infty$), and make predictions for future experiments. First, we use a time-dependent variational ansatz in a many-body theory with {\it mean-field-like} approximations and a regularized effective potential, expecting the results to be valid at short times when the condensate depletion is small.
We then compare the results of this many-body calculation to an exactly solvable two-body model~\cite{BuschTwoColdAtoms}. We argue that, when the system undergoes a sudden quench in the scattering length, the early stages of time evolution correspond to the build up of local equilibrium between nearby particles, see Fig.~\ref{fig-cartoon}.  The earliest stages of this evolution, therefore, simply involve the dynamics of a two-body system; a similar rationale was employed for weaker interactions in Refs.~\cite{GreeneArtificialTrapsForTwoBodySystemsNJP2003,PaulJulienneTwoBodyArtificialTrapsJPhysB2004}).
Remarkably, this seemingly naive argument turns out to provide an intuitive understanding as well as strong quantitative agreement with the many-body model at large momentum and short times compared to the scales set by the density.
This is perhaps less surprising in light of the connection between Tan's contact (which determines the occupation of large momentum modes) and two-body collision physics~\cite{Tan123,ZhangLeggettPRA2009}.
We also discuss the results of a three-body model that is similar in spirit to the two-body model, but which assumes a Lennard-Jones potential for the interparticle interactions. We compare the timescales for which correlation dynamics occurs  to the timescale for three-body loss, and find results which are consistent with time scales reported in Ref.~\cite{PhilDebbieEricUnitaryBoseGas}. Finally, we make predictions for future experiments regarding the structure of the momentum distribution, especially the contact dynamics.

{\it Many-body variational calculation:} We consider the many-body Hamiltonian
\begin{equation}
\hat H = \sum_{\bf k}\epsilon_{\bf k}\hat a_{\bf k}^{\dagger}\hat a_{\bf k}+\frac{1}{2V}\sum_{{\bf k}_1,{\bf k}_2,{\bf q}}\tilde{U}({\bf q})\hat a_{{\bf k}_1+{\bf q}}^{\dagger}\hat a_{{\bf k}_2-{\bf q}}^{\dagger}\hat a_{{\bf k}_1}\hat a_{{\bf k}_2},
\end{equation}
where $V$ is the system volume, $\tilde{U}({\bf q})=\int d^3\! {\bf r}\, \mathrm{e}^{-i{\bf q}\cdot {\bf r}}U({\bf r})$ is the Fourier transform of the interaction potential, $\epsilon_{{\bf k}}=\hbar^2k^2/2m$ is the single-particle kinetic energy, and $\hat a_{\bf k}$ ($\hat a_{\bf k}^{\dagger}$) is a bosonic annihilation (creation) operator for a particle of momentum $\hbar {\bf k}$. We use a time-dependent generalization of the ansatz introduced in Ref.~\cite{SongZhouPRL}:
\begin{equation}\label{eq-ManyBodyAnsatz}
| \Psi_{\rm var}(t) \rangle = \mathcal{A}(t) \mathrm{exp}\left[ c_0(t) \hat a_0^{\dagger}+\sum_{{\bf k}\cdot \hat{\bf z}>0} g_{\bf k} (t)\hat a_{\bf k}^{\dagger}\hat a_{-\bf k}^{\dagger}\right] |0\rangle ,
\end{equation}
where $\mathcal{A}(t)$ is a normalization constant that depends on the variational parameters $\{ c_0(t),g_{{\bf k}\neq 0}(t) \}$, and $| 0 \rangle$ is the particle vacuum. At short times before condensate depletion becomes large this ansatz is justified by the Bogoliubov-like idea that the condensed mode behaves as a coherent state and the excited particles are generated in pairs by the dominant term $\hat a_{\bf k}^{\dagger}\hat a_{-\bf k}^{\dagger}\hat a_0 \hat a_0$ in $\hat H$. Our variational parameters are related to momentum occupations via $n_0(t) = |c_0(t)|^2$ and $n_{\bf k}(t)=|g_{\bf k}(t)|^2/(1-|g_{\bf k}(t)|^2)$.  We note that average total particle number and energy are conserved.

In terms of these amplitudes, the equations of motion for the system are $i\hbar \dot{c}_0=\partial\langle\hat H\rangle/\partial c_0^*$ and $i\hbar \dot g_{\bf k}=(1-|g_{\bf k}|^2)\partial\langle\hat H\rangle/\partial g_{\bf k}^*$. We solve these coupled equations numerically under the assumption that $g_{\bf k}$ depends only on $\left|{\bf k}\right|$, and our initial condition is chosen to be a noninteracting gas such that $g_{\bf k}(0)=0$ for all ${\bf k}\neq 0$. The short-range interactions are modeled with an attractive spherical square well $U({\bf r})mr_0^2/\hbar^2=-\left(\frac{\pi}{2}\right)^2\Theta(r_0-r)$, where $r_0$ is the range of the potential and is assumed to be much smaller than the interparticle spacing. The depth of the well is chosen such that there is a single two-body bound state at threshold, so the scattering length diverges~\cite{AstrakharchikPRL93-200404-2004}. We have found that the dynamics for low momenta ($kr_0\ll 1$) do not depend on the specific choice of $r_0$ as long as $nr_0^3\ll 1$; additionally, the computed dynamics are found to scale universally with the appropriate density units. We discuss the results of this model after introducing a complementary two-body model that yields comparable results.

{\it Two body model:}
Based on the idea of dynamics originating in causally isolated regions of the gas (see Fig.~\ref{fig-cartoon}),
 we consider the dynamics of a two-body wavefunction in an artificial trap, $\psi_{\rm trap}^{\rm (2B)}({\bf r};t)\!=\!\sum_\nu\! c_\nu\psi_{\nu}({\bf r})e^{-iE_\nu t/\hbar}$, with ${\bf r}={\bf r}_1-{\bf r}_2$ being the relative coordinate. The summation is over all eigenvalues/states~\cite{BuschTwoColdAtoms} of the Hamiltonian;
\begin{equation}\label{EQ-hamiltonian}
 \left[-\frac{\hbar^2}{2\mu}\nabla^2_{{\bf r}}+\frac{1}{2}\mu\omega_{\rm ho} r^2+
\frac{2\pi\hbar^2a}{\mu}
\delta^{(3)}_{\rm reg}({\bf r})\right]\psi_{\nu}({\bf r})=E_\nu\psi_{\nu}({\bf r}),
\end{equation}
with $\mu=m/2$ being the reduced mass, and $\delta^{(3)}_{\rm reg}({\bf r})=\delta^{(3)}({\bf r})\partial_rr$ is the Fermi pseudo-potential.
The coefficients $c_\nu=\int d^3{\bf r}\;\psi_{\rm trap}^{\rm (2B)}({\bf r};0)\psi_\nu^*({\bf r}) $ are determined by the initial condition, which we choose to be the noninteracting ground state $\psi_{\rm trap}^{\rm (2B)}({\bf r};0)=\pi^{-3/4}a_{\rm ho}^{-3/2}e^{-r^2/2a_{\rm ho}^2}$, where $a_{\rm ho}=\sqrt{\hbar/\mu\omega_{\rm ho}}$ is the harmonic oscillator length.
The trap width, $a_{\rm ho}$, is the only free parameter in the two-body model. We choose $a_{\rm ho}$ such that the average separation of the two particles is $\langle r\rangle=\int d^3 {\bf r}\;r|\psi_{\rm trap}^{\rm (2B)}({\bf r};0)|^2=\frac{2a_{\rm ho}}{\sqrt{\pi}}\defeq(4\pi n/3)^{-1/3}$, where $n$ is the initial density of the actual many-body system. Thus, the two particles initially have a mean interparticle separation that is consistent with the many-body system. We expect this two-body model to be relevant on time scales much less than the trap period $2\pi/\omega_{\rm ho}$.

At unitarity, the eigenvalues of Eq.~\eqref{EQ-hamiltonian} are  $E_\nu=(2\nu+3/2)\hbar\omega_{\rm ho}$, with $\nu\in\left\{-1/2,1/2,3/2,\ldots\right\}$, and the normalized eigenstates~\cite{BuschTwoColdAtoms}  are $ \psi_{\nu}({\bf r})=A_\nu \frac{1}{\tilde{r}}e^{-\tilde{r}^2/2}{\rm H}_{2\nu+1}(\tilde{r})$.
Here ${\rm H}_{n}$ is a Hermite polynomial, $\tilde{r}=r/a_{\rm ho}$, and $A_\nu\!=\!\frac{2^{-1-\nu}}{\pi (2\nu)!!a_{\rm ho}^{3/2}}\sqrt{\frac{\Gamma(1+\nu)}{\Gamma(3/2+\nu)}}$.
We compute the overlap integrals to find $c_{\nu}\!=\!-\frac{(-2)^{\nu+1/2}\pi^{1/4}(2\nu)!!}{\nu}a_{\rm ho}^{3/2}A_\nu$.
With the momentum-space eigenstates $\tilde{\psi}_\nu({\bf k})$~\cite{EPAPSarticle}, it is straightforward to calculate the momentum distribution $n^{\rm (2B)}_k(t)=(2\pi)^{-3}|\sum_{\nu} c_\nu\tilde{\psi}_\nu({\bf k})e^{-iE_{\nu}t/\hbar}|^2$. We assume that the center of mass of the two atoms is untrapped, in which case the total momentum distribution coincides with the relative momentum distribution.

For $k$ much greater than the effective trapping length  and times $t$ much smaller than the trap period, the momentum distribution dynamics reduces to the simple analytic formula
\begin{align}\label{eq-KadenKoller}
n_{k}(t) &=
\frac{An}{k^4}\left|\frac{2j}{\sqrt{\pi}}\sqrt{t \mu/\hbar}e^{i \frac{\tau}{2}}-\frac{1}{k} \text{erf}\left( j\sqrt{\tau}\right)\right|^2
\end{align}
where $j=e^{-i\pi/4}/\sqrt{2}$, and $\tau=t/\tau_k$ with $\tau_k=\mu/(\hbar k^2)$ denoting a characteristic time-scale for each momentum mode.  This formula is independent of the trapping potential except through the prefactor $A$, which we have explicitly verified (see Ref.~\cite{EPAPSarticle}). We note that for $^{85}$Rb at a density of $5.5\times10^{12}$cm$^{-3}$, we have $\tau_{k_{\rm F}}\approx20\mu$s, $\tau_{2k_{\rm F}}\approx6\mu$s, and  $\tau_{3k_{\rm F}}\approx2.5\mu$s, which is comparable to the timescales reported in Fig. 4 of  Ref.~\cite{PhilDebbieEricUnitaryBoseGas} (although a direct comparison is not entirely appropriate).

\begin{figure}[t!]
 \includegraphics[width=8cm]{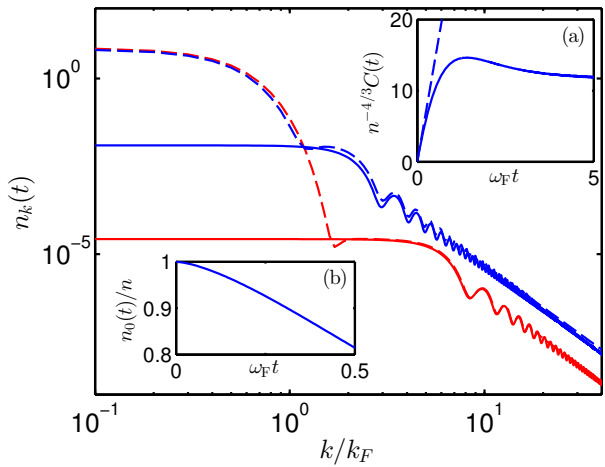}
\caption{(Color online) The momentum distribution for two different times $\omega_{\rm F}t=0.05$ (shown in red) and $\omega_{\rm F}t=0.4$ (shown in blue). The solid lines show results from the many-body model, and the dashed lines show results from the harmonically trapped two-body model. Note that $n_k=n(2\pi)^3n_k^{\rm (2B)}$~\cite{EPAPSarticle}. Inset (a) shows the time evolution of the contact (dashed line shows Eq.~\eqref{eq-contact}), which saturates beyond $\omega_{\rm F}t\gtrsim 3$.   Inset (b) shows the evolving condensate fraction of the many-body model.}
\label{fig-FixedTimePlot}
\end{figure}

{\it Results:} Figure~\ref{fig-FixedTimePlot} shows plots of the momentum distribution for two rescaled times  $\omega_{\rm F}t$, where $\omega_{\rm F}=(6\pi^2n)^{2/3}\hbar/2m$ is the Fermi frequency~\cite{footnote}. We have found that the depletion of the condensate remains below $15\%$ for $\omega_{\rm F}t\lesssim 0.4$ [inset (b)], which establishes an approximate regime of validity for the many-body ansatz, Eq.~\eqref{eq-ManyBodyAnsatz}.
The agreement between two- and many-body models at high momenta, $k_{\rm F}\ll k\ll 1/r_0$, is excellent, whereas the discrepancy at low momenta, $k\lesssim k_{\rm F}$, is due to the initial condition $\psi_{\rm trap}^{\rm (2B)}({\bf r};0)$ of the two-body model (dashed lines in Fig.~\ref{fig-FixedTimePlot}).
At high momenta, both calculations reveal a dependence $n_k\sim1/k^4$, with oscillations.  This behavior is characteristic of short-range two-body interactions and leads to the thermodynamic parameter known as the contact, $C\equiv \lim_{k\to \infty}k^4n_{k}$ (normalized such that $\sum_{\bf k}n_{\bf k}=N$)~\cite{Tan123}.  Our time-dependent simulations show that the $1/k^4$ tail first appears at large momenta  and then propagates in to smaller momenta.
The contact initially exhibits linear growth, which can be extracted from the asymptotics of the two-body model~\cite{EPAPSarticle}:
\begin{equation}\label{eq-contact}
 C(t)=2048\left(\frac{2n^4}{3^5\pi^7}\right)^{1/3}\omega_{\rm F}t.
\end{equation}
This formula agrees (within $6\%$) with a linear fit to the many-body numerical data (see the inset of Fig.~\ref{fig-FixedTimePlot}), which predicts $C(t)=26.9(1)n^{4/3}\omega_{\rm F}t$ for $\omega_{\rm F}t\lesssim 0.2$.
Recall, however, that deriving Eq.~\eqref{eq-contact} involves fixing the free parameter in the two-body model such that $\langle r \rangle=(4\pi n/3)^{-1/3}$. There exist alternative estimates for the interparticle spacing, and these could change the prefactor in Eq.~\eqref{eq-contact} by up to an order of magnitude.
\begin{figure}
 \includegraphics[width=8cm]{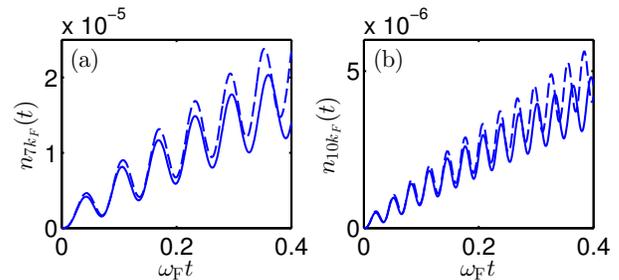}
\caption{(Color online) The occupation of a single (fixed) momentum mode as a function of time. In (a) $k=7k_{\rm F}$, and in (b) $k=10k_{\rm F}$. 
To compare with Ref.~\cite{PhilDebbieEricUnitaryBoseGas} we note that $\omega_{\rm F}t=0.4$ corresponds to approximately 34$\mu$s for $^{85}$Rb at a density of $5.5\times10^{12}{\rm cm}^{-3}$.}
\label{fig-FixedMomentumPlot}
\end{figure}

The two models differ markedly at longer times. The two-body model is completely periodic with the period of the artificial trap, $T_{\rm ho}=\frac{2\pi}{\omega_{\rm ho}}=\frac{3\pi^{8/3}}{1\!6\,\omega_{\rm F}}\left(\frac{3}{2}\right)^{1/3}$. The many-body model, however, predicts a relaxation of the contact into an asymptotic value of $C\approx 12n^{4/3}$ [Fig.~\ref{fig-FixedTimePlot}, inset (a) ], remarkably close to the unitary ground-state prediction of Ref.~\cite{StoofJastrowNumerics}, but different from that of~\cite{Stoof1UnitaryBoseGasRGTheory2013,Stoof2UnitaryBoseGasRGTheory2013} ($C\approx32n^{4/3}$). We note that our prediction for the contact at large times is not rigorously justified because the accompanying depletion (which eventually approaches  $\sim 100\%$) is large enough to violate assumptions of the ansatz, Eq.~\eqref{eq-ManyBodyAnsatz}. The two-body model also predicts a certain probability of molecule formation when the quench is ultimately reversed back to a small scattering length. Depending on the time spent at unitarity, our calculations predict fractions up to about 20\%.

Figure~\ref{fig-FixedMomentumPlot} shows the time evolution of a single momentum mode in the system, for two different momenta, both well above $k_F$. The agreement between the two- and many-body theories is excellent at short times, suggesting that the theoretical approach of isolating two particles in this way is reasonable in this regime.  Overall, the variation of $n_k$ with time exhibits linear growth consistent with Eqs.~\eqref{eq-KadenKoller} and \eqref{eq-contact}, plus oscillations (similar to the weakly interacting limit~\cite{NatuMueller}) on the time scale $\tau_k$ defined below Eq.~\eqref{eq-KadenKoller}.  These oscillations were not observed in the experiment~\cite{PhilDebbieEricUnitaryBoseGas}. We surmise that this is due either to the experimental averaging over the nonuniform density profile, thermal initial states, or many-body correlations omitted from our theory.

{\it Three body model:}
One of the crucial results of Ref.~\cite{PhilDebbieEricUnitaryBoseGas} is the long time scales for three-body loss relative to the observed dynamics at unitarity. With a goal of understanding this separation of timescales, we now analyse the three-body physics relevant for the experiment in Ref.~\cite{PhilDebbieEricUnitaryBoseGas} and explore the mechanisms behind the relative stability against three-body losses.

For values of $k|a|\gg 1$, it is well known that three-body loss, described by the rate equation $\dot{n}=-L_{3} n^3$, is independent of $a$
\cite{DIncaoPRL2004,SalomonLifetimeWithResonantInteractionsFiniteTemp} and can be approximately characterized at a given temperature $T$ by
\begin{eqnarray}
L_{3}(k_BT)=\frac{36\sqrt{3}\pi^2}{m^3}\frac{(1-e^{-4\eta})}{(k_{B}T)^2}\hbar^5, \label{L3Unit}
\end{eqnarray}
where $\eta>0$ is the inelasticity parameter~\cite{SalomonLifetimeWithResonantInteractionsFiniteTemp} (accounting for decay to deeply bound molecular states).
Using Eq.~(\ref{L3Unit}) for the $^{85}$Rb experiment ($T\approx10$nK \cite{PhilDebbieEricUnitaryBoseGas}, $\eta\approx 0.06$ \cite{85Rb_JILA}, and
peak-density $n_{pk}\approx9.6\times10^{12}$cm$^{-3}$), one obtains a lifetime estimate, $\tau=1/(n_{pk}^2L_{3})$, of about 0.4$\mu$s, i.e., much shorter than the experimentally observed value $\tau_{exp}\approx0.63$ms.
However, for a gas at such low temperatures, the thermal de Broglie wavelength is ten times larger than the mean interparticle spacing and the validity of Eq.~\eqref{L3Unit} comes into question. Based purely on arguments of dimensional analysis, one can substitute $k_BT$ in Eq.~\eqref{L3Unit} with the Fermi energy
$E_{\rm F}=\hbar\omega_{\rm F}$
\cite{PhilDebbieEricUnitaryBoseGas}. In doing so, the rate constant becomes density dependent so we also need to average over the cloud accordingly;
$\langle L_3\rangle=\int L_3 n^3 d^3{\bf r}/\int n^3d^3{\bf r}$. Assuming a Thomas-Fermi distribution 
the average recombination rate is given by:
\begin{eqnarray}
\langle{L_{3}}\rangle=\frac{945~\Gamma(8/3)}{8~2^{1/3}~\Gamma(25/6)}\left(\frac{3}{\pi}\right)^{1/6}\frac{(1-e^{-4\eta})}{n_{pk}^{4/3}}\frac{\hbar}{m},
\label{L3Avg}
\end{eqnarray}
which leads to
$\langle\tau\rangle=21/8n^2_{pk}\langle L_{3}\rangle\approx0.20$ms, where the angle brackets imply a spatial average ~\cite{EPAPSarticle}.
This result is in better agreement with the observed value. However we stress that this approach is derived from the concept of scattering (three asymptotically free particles) which
loses meaning at unitarity ($na^3\gg1$).
 
To improve upon the dimensional-analysis argument of the previous paragraph, we numerically solve the three-body problem
\cite{SunoPRA2002,WangPRL2012} in a harmonic trap whose confinement length, $a_{\rm ho}$,
is set by the average interatomic distance as with our two-body model. For the three-body calculations,
the value of $a_{\rm ho}$ is chosen slightly differently to retain the correct mean-interparticle-spacing.
In our model, three-body states have a finite width (due to decay into deeply bound states,
included via an artificial channel adjusted to reproduce $\eta\approx0.06$ for $^{85}$Rb~\cite{85Rb_JILA}).
Importantly, this approach to the three-body problem does not rely on the concept of scattering.
The average {\em effective} recombination rate can be defined as,
\begin{eqnarray}
\langle{L_{3}^{\ast}}\rangle=\frac{21}{8N_0}\frac{1}{n_{pk}^2}\sum_{\beta}\int\left[c_{\beta}^2(a_{\rm ho})\frac{\Gamma_{\beta}(a_{\rm ho})}{\hbar}\right]n(r)d^3{\bf r}.\label{L3AvgNum}
\end{eqnarray}
Here, $c_{\beta}$ is the probability of populating the three-body state $\beta$ after instantaneously switching $a$ from $a\approx150a_0$ ($a_0$ being the Bohr radius) to $a=\infty$, $\Gamma_{\beta}$ is the width of the corresponding state, and $N_0$ is the total number of particles ($7\times10^4$ in the case of Ref.~\cite{PhilDebbieEricUnitaryBoseGas}).
Note that in the above equation both $c_{\beta}$ and $\Gamma_{\beta}$ depend on $r$ via the dependence of $a_{\rm ho}$ on the density (which is assumed to be a Thomas-Fermi profile).
The results lead to an average lifetime of
$\langle{\tau^{\ast}}\rangle\approx1.1$ms, also consistent with the experimental observations.
Therefore, our analysis on the lifetime due to three-body losses at unitarity allows us to conclude
that the longer lifetimes observed in Ref.~\cite{PhilDebbieEricUnitaryBoseGas} are likely a general effect that can open up ways to explore strongly correlated Bose gases. This numerical result produces the same $n^{2/3}$ scaling of the loss rate $\gamma = n^2 L_3$ as does the simpler formula (7).  In addition, this loss exhibits weak ($\sim 20\%$) oscillations over several decades of density due to Efimov effects, a topic left to future study.
 
In summary, we have explored the short-time evolution of a degenerate Bose gas upon quenching to an infinite scattering length.  We introduced and solved few- and many-body models and demonstrated the role of two-body physics in the subsequent dynamics.  Specifically, we characterized the evolution of the momentum distribution on time scales shorter than the characteristic {\it Fermi time}. We calculated the oscillation-timescale associated with each momentum mode and found that it scales as $1/k^2$, demonstrating how populations of higher momenta approach $1/k^4$ more quickly than those of lower momenta. We observed the emergence of Tan's contact in our simulations, and made predictions for its growth at short times [see Eq.~\eqref{eq-contact} and surrounding text].  Following on with the idea of locality in the interactions, we also make predictions for three-body losses, and found that the characteristic time scale of this loss is longer than that of the two-body dynamics. These findings are qualitatively consistent with experimental results~\cite{PhilDebbieEricUnitaryBoseGas}, and provide theoretical support to the idea that the observation of a locally equilibrated unitary Bose gas may indeed be possible.  It would be fascinating to test our predictions in future experiments, perhaps with scattering or spectroscopy.
In closing, we remark that the final (quasi-) equilibrium state of the system following such a quench is beyond the scope of our current theory. At longer times, effects due to strong correlations between larger numbers of particles presumably become important. The task of understanding the nature of these higher-order correlations, and at which timescales they present themselves, remains for future work.

We acknowledge Stefan Natu, Erich Mueller, Catherine Klaus, Philip Makotyn, Debbie Jin, and Eric Cornell for suggestions.  We gratefully acknowledge funding from the NSF (J.L.B.), ARO MURI (A.G.S.), NDSEG fellowship program (J.P.C. and A.P.K.), AFOSR MURI and NSF (J.P.D. and C.H.G.),  and  NSF-PFC  (K.R.A.H. and A.M.R.).  K.R.A.H. thanks the NRC for support.  This manuscript is the contribution of NIST and is not subject to U.~S. copyright.

\newpage

{\bf Supplementary material for: Quenching to unitarity: Quantum dynamics in a 3D Bose gas}


{\it Two-body model:} We find an analytic formula for the Fourier transform of the eigenstates of two interacting atoms in a harmonic trap (following Reference [22] of the main text) at unitarity,
\begin{align}
 \tilde{\psi}_\nu({\bf k})=&\frac{4\pi A_\nu }{\tilde{k}}
\left[(-2)^{\nu+3/2}{\rm Q}_{2\nu+1}(\tilde{k})+\phantom{\left(\frac{\tilde{k}}{\sqrt{2}}\right)}\right.\nonumber\\
&
\left.
\sqrt{2}(-1)^{\nu+1/2}
\fd\left(\frac{\tilde{k}}{\sqrt{2}}\right)
\herm_{2\nu+1}(\tilde{k})\right]
\end{align}
where $\tilde{k}=ka_{\rm ho}$ is a dimensionless momentum, $\fd$ is the Dawson integral, and
\begin{align}
 Q_m(k)=&m!\sum_{l=0}^{m/2}\frac{(-1)^{m/2-l}2^{2l}}{(2l)!(m/2-l)!}\times\nonumber\\
&
\sum_{n={\rm max}(0,\ceil{\frac{2l-m-1}{2}})}
^{\floor{l-1/2}}(2n-1)!!\,k^{2l-2n-1}
\end{align}
is a polynomial ($\floor{\cdot}$ and $\ceil{\cdot}$ indicate the floor and ceiling functions respectively).

To extract the behaviour of the contact from our results, we find the asymptotic form of $\tilde{\psi}_\nu({\bf k})$ as $k\rightarrow\infty$,
\begin{equation}
 \tilde{\psi}_\nu({\bf k})\rightarrow a_{\rm ho}^{3/2} (-1)^{\nu+1/2}2^{1-\nu}\pi^{1/4}\sqrt{\frac{\Gamma\left(2+2\nu\right)}{\Gamma\left(3/2+\nu\right)}}\frac{1}{\tilde{k}^2}.
\end{equation}
Using this formula, along with the expression for the overlap integral $c_\nu$ given in the main text, we find
\begin{equation}\label{eq-nk}
 n_{k\rightarrow\infty}^{\rm (2B)}=\frac{4a_{\rm ho}^3}{\pi^{7/2}\tilde{k}^4}|\sin(\omega_{\rm ho}t)|,
\end{equation}
and since we only trust the correspondence between the model and the physical system at times small compared to the harmonic oscillator time, we Taylor expand to get $n_{k\rightarrow\infty}^{\rm (2B)}=\frac{4a_{\rm ho}^3}{\pi^{7/2}\tilde{k}^4}\omega_{\rm ho}t$.

Finally, we note that $n_{k}^{\rm (2B)}$ is normalised such that $\int d^3{\bf k}\, n_{k}^{\rm (2B)}=1$, as opposed to the usual many-body normalization where $n_{k}$ is dimensionless and $\int \frac{d^3{\bf k}}{(2\pi)^3} \,n_{k}=n$. Multiplying Eq.~\eqref{eq-nk} by $(2\pi)^3n$, and using the relationship between $a_{\rm ho}$ and $n$ proposed in the main text, yields Eq. (5) of the main text.

{\it Large-momentum, short time two-body dynamics:}  As described in the main text, the key features of the dynamics of the two-body model are captured by a simple, analytic formula [Eq.~(4), main text] that is valid for momenta large compared to the inverse of the length scale of the confining potential.  This formula approximately describes the time dependence of the large-momentum, many-body results as well.

Equation~(4) in the main text captures the characteristic fast oscillations at momentum $k$ with timescale $\tau_k=\mu/(\hbar k^2)$, and the linear rise in the $1/k^4$ tail.  As argued in the main text, up to a constant prefactor the dynamics in this large-momentum, short-time limit result should be independent of the details of the effective trapping potential used.  We have explicitly verified this by comparing calculations for a spherical box with hard walls, the harmonic oscillator model introduced in the main text, and a continuum model described below.

We briefly sketch the derivation.  We work in a continuum, and will regularize all wavefunctions by assuming a decay $e^{-r/(2L)}$, with $L$ large compared to other scales in the problem; such regularization is implicit in the equations below. We assume the system starts in a non-interacting wavefunction with momentum $k_0$ in relative spherical coordinates
\begin{equation}
\phi_{k_0}(r) = A_{k_0} \sin(k_0 r)/r
\end{equation}
where the normalization $A_{k_0}$ matters only for the prefactor.  The eigenstates for atoms interacting with a pseudopotential at unitarity are
\begin{equation}
\varphi_{p}(r) = B_{p} \cos(p r)/r
\end{equation}
where $B_p$ is a normalization factor.  The eigenenergies are $\hbar^2k^2/(2\mu)$. The time dynamics of the two-body wavefunction is then given by
\begin{equation}
\phi_{k_0}(r,t) = \int_0^\infty \!dp\, c_{k_0 p}\varphi_{p}(r) e^{-i\hbar^2 k^2 t/(2\mu)},
\end{equation}
where $c_{k_0 p}$ is the projection of $\phi_{k_0}$ on $\varphi_{p}$.  Calculating these coefficients and the integral over $p$, and from this calculating the momentum distribution  $n_k = (2\pi)^3 n |\int \! d^3r\, e^{i\vec{k}\cdot\vec{r}} \phi_{k_0}(r,t)|^2$,
we find the formula given by Eq.~(4) in the main text.

{\it Three-body model:}
The three-body loss rate, $L_{3}$, is determined from the rate equation
$\dot{n}=-L_{3}n^3$, after integrating over the entire desity distribution:
\begin{eqnarray}
\int \dot{n}(\vec{r},t)d^3r=-\int L_{3}(r) n^3(\vec{r},t) d^3r.
\end{eqnarray}
Although the position-dependent rate $L_3(r)$  is unobservable in current experiments,  one accessible observable is the {\em averaged} recombination rate, $\langle L_{3}\rangle$. Assuming a Thomas-Fermi distribution
$n(r)=n_{pk}\left[1-(r/r_{TF})^2\right]$, where $n_{pk}$ is the peak density and $r_{TF}$ the Thomas-Fermi radius, we find
\begin{eqnarray}
\langle L_{3}\rangle &=& {\int L_{3}(r) n^3(\vec{r},t) d^3r}\Big/{\int n^3(\vec{r},t) d^3r} \nonumber\\
&=& \frac{21}{8N_{0}}\frac{1}{n_{pk}^2}{\int L_{3}(r) n^3(\vec{r},t) d^3r},
\end{eqnarray}
where $N_{0}$ is the initial atom number. The above expression leads to an approximate rate equation
\begin{eqnarray}
\dot{N}(t) \approx-\langle L_3\rangle \left(\frac{8N_0^2}{21n_{pk}^2}\right) N^3(t),\label{LossRate}
\end{eqnarray}
which is expected to be valid at short times  where $N(t)\approx N_0$. As experiments frequently fit to time-evolution of the atom number $N(t)$ given by Eq.~\eqref{LossRate}, we base our analysis of three-body losses on this.  In particular, we use  the {\em average} loss rate $\langle \Gamma\rangle$ and {\em average}
lifetime $\langle \tau\rangle$ as
\begin{eqnarray}
\langle \Gamma\rangle=\left(\frac{8 n_{pk}^2}{21N_{0}^2}\right) \langle L_{3}\rangle~~~\mbox{and}~~~
\langle \tau\rangle=\frac{1}{N_0^2 \langle \Gamma\rangle},
\end{eqnarray}
respectively, to compare with experiment.


\begin{thebibliography}{99}
\bibitem{footnote}The Fermi momentum/energy simply give us convenient units when the interparticle spacing is the only length scale, no notion of Fermionization (as in that which occurs in strongly interacting 1D bosons) is implied by using such units.
\end{thebibliography}
\end{document}